\input{aipcheck}

\documentclass[
    final            
  ,numberedheadings, numcites 
  ]
  {aipproc}
\usepackage{epsf}

\layoutstyle{8d}

\def\beq{\begin{equation}}
\def\eeq{\end{equation}}
\def\bea{\begin{eqnarray}}
\def\eea{\end{eqnarray}}
\def\beqa{\begin{equation}\begin{array}{l}}
\def\eeqa{\end{array}\end{equation}}
\def\eqlab#1{\label{eq:#1}}
\def\figlab#1{\label{fig:#1}}

\def\Eqref#1{Eq.~(\ref{eq:#1})}

\def\Figref#1{Fig.~\ref{fig:#1}}

\def\al{\alpha}
\def\be{\beta}
\def\ga{\gamma} 
\def\de{\delta} \def\De{\Delta}

\def\si{\sigma}

\def\nn{\nonumber}

\begin{document}

\title{Nucleon polarizabilities and $\Delta$-resonance magnetic moment in
 chiral EFT }

\classification{14.20.Dh, 14.20.Gk, 13.60.Fz, 13.60.Le, 11.30.Rd, 11.55.Fv}
 
\keywords      {chiral expansion, dispersion relations, electromagnetic polarizabilties, resonance magnetic moment}

\author{Vladimir Pascalutsa}{
  address={Institut f\"ur Kernphysik, Johannes Gutenberg Universit\"at, Mainz D-55099, Germany}
}

\begin{abstract}
 Recent chiral EFT calculations of nucleon polarizabilities 
reveal a problem in the current empirical determination of  proton's 
electromagnetic polarizabilities. We also report on the progress
in the empirical determination of the $\Delta$(1232)-resonance magnetic moment
in the process of $\gamma p \to p  \pi^0 \gamma'$ measured at MAMI.
\end{abstract}

\maketitle


\section{Introduction}
The chiral effective-field theory ($\chi$EFT)  describes the physics of low-energy strong interaction
in terms of hadronic degrees of freedom, rather than in terms of the underlying quarks and gluons.
Originally $\chi$EFT dealt with only the Nambu-Goldstone bosons (pions, kaons) of the spontaneous chiral symmetry breaking \cite{Weinberg:1978kz, Gasser:1983yg}, 
but the lightest baryons have eventually been
included too~\cite{GSS89}. 
The inclusion of baryons has proved to be difficult and even until now there is
no  consensus on how to count the baryon contributions. However, my aim here is not
to discuss the various counting schemes; this is done elsewhere~\cite{Pas10}. 
I am concerned here with the question of what  $\chi$EFT, combined
with experiment, can tell us about the nucleon and its first excitation, the $\Delta$(1232)-isobar.

The two sets of quantities considered here (nucleon polarizabilities and $\Delta$'s
magnetic dipole moment) bear at least one thing in common: 
they are not measured directly but rather are extracted
 from experimental data using theoretical modeling.
In principle the $\chi$EFT should be a suitable framework
for this task, and we are going to focus on it here. 
 The results for polarizabilities have been obtained
in collaboration with Vadim Lensky \cite{Lensky:2008re,Lensky:2009uv}, while the work on  $\Delta$'s MDM is being
done together with Marc Vanderhaeghen \cite{Pascalutsa:2004je,Pascalutsa:2007wb}. In both cases I will present here some
new, thusfar unpublished results.


\section{Compton scattering: Nucleon polarizabilities}
Proton's electric ($\alpha$) and magnetic ($\beta$) polarizabilities
are determined in low-energy Compton scattering ($\ga p \to \ga p$). The purely
elastic process is described by Born contributions which give rise
to the Powell cross section. Polarizabilities
are characterized by the inelastic contribution. At the level of unpolarized
differential cross section we have:
\bea
\eqlab{unpolcs}
&& \frac{d\si}{d\Omega} - \frac{d\si}{d\Omega}^{\!\!(\mathrm{Born})}    =  
- \frac{e^2}{8\pi M_p}\left( \frac{\nu'}{\nu} \right)^2\,  \nu\nu'   \\
&& \quad \times \,  \Big[  (\alpha+\be) (1+z)^2   +(\al- \beta)(1- z)^2  \Big] + O(\nu^3), \nn
\eea
where $e$ and $M_p$ are the proton charge and mass, $\nu$ ($\nu'$) is the  incoming (outgoing) photon energy in the lab system, 
$z$ is the cosine of the lab scattering angle, $d\Omega = 2\pi \,dz$. This is a model-independent
result and the Born term is known, hence by measuring the angular distribution
at very low energy one should be able to determine $\al$ and $\be$. 

However, the experiments
at low energies are very tough. So far most of the data are obtained for energies above 100 MeV,
where the low-energy expansion begins to diverge, due to proximity of the pion production cut which sets the convergence radius. The last and the best ---in terms of statistics and coverage--- experiment, performed by the A2/TAPS Collaboration at MAMI \cite{MAMI01}, employed a dispersion-relation
(DR) model of L'vov \cite{Lvov:1980wp} to extract $\al$ and $\be$. The result of this extraction
is displayed in \Figref{potato} by a dashed ellipse labeled `TAPS'. Applying the Baldin sum rule
value for the sum of polarizabilities cuts out the `global average' result which is quoted
by the Particle Data Group \cite{PDG2010}, namely :
\bea
\eqlab{PDGvalues} 
\hskip-1cm\mbox{PDG:} && \al = (12.0\pm 0.6) \times 10^{-4} \mbox{ fm$^3$} ,\nn\\
&& \be = (1.9\pm 0.5) \times 10^{-4} \mbox{ fm$^3$}\,.
\eea

\begin{figure}[tb]
\centerline{\epsfclipon  
\epsfxsize=7cm%
  \epsffile{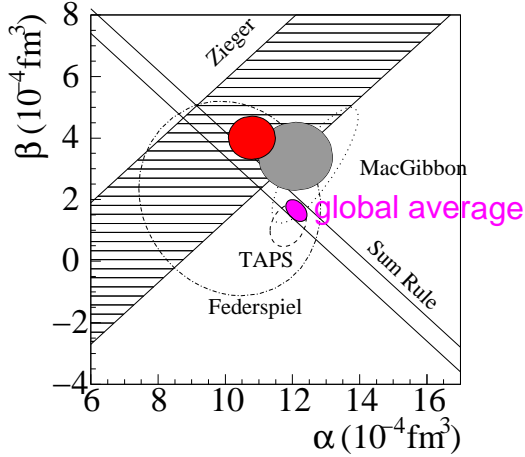} 
}
\caption{The NNLO BChPT result \cite{Lensky:2009uv} is shown by the red blob and the $\De$-less  HB$\chi$PT
result~\cite{Beane:2004ra} is shown by the grey blob. `Sum Rule' indicates the Baldin sum rule constraint on $\alpha+\beta$. Experimental
results are from Federspiel et~al.~\cite{Federspiel:1991yd},
Zieger et al.~\cite{Zieger:1992jq}, MacGibbon et al.~\cite{MacG95},
and TAPS~\cite{MAMI01}.
`Global average' represents the PDG summary~\cite{PDG2010}.} 
\figlab{potato}
\end{figure}

\begin{figure}
 \centerline{\epsfclipon  \epsfxsize=6cm%
  \epsffile{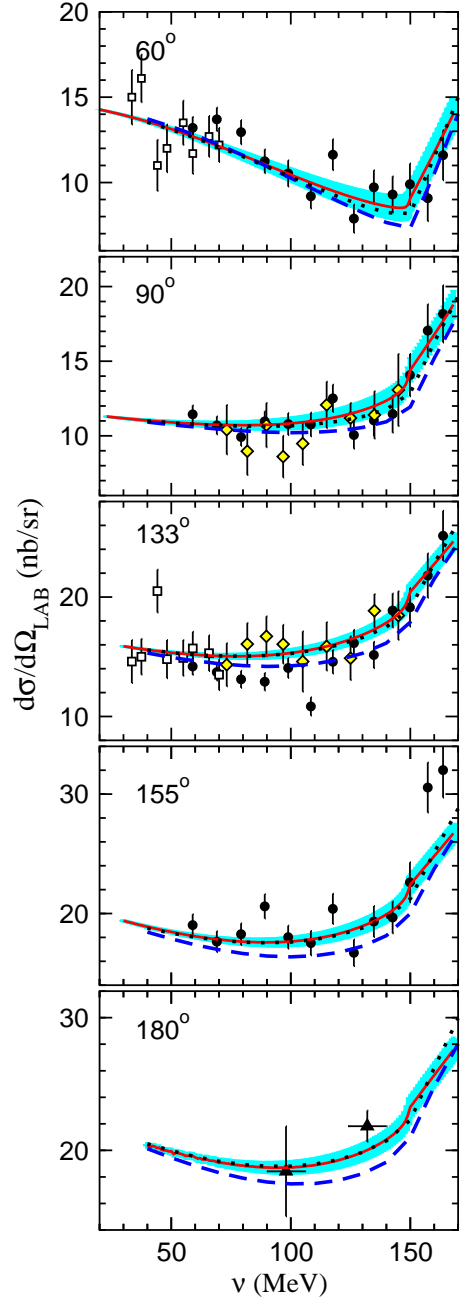} 
}
  \caption{\small  Energy dependence of $\ga p \to \ga p$ differential cross section, at fixed scattering angles. The curves are explained in the text.}
\figlab{DRvsEFT}
\end{figure}

\bigskip 
What is remarkable is that the present $\chi$EFT calculations
 \cite{Lensky:2009uv,Beane:2004ra} differ from PDG
 in the values for polarizabilities (see the red and the
 grey blobs) while agreeing with the cross-section data. For instance,
 the next-to-next-to-leading order (NNLO) calculation in baryon chiral perturbation theory (B$\chi$PT)
yields  \cite{Lensky:2009uv}: 
\bea
\eqlab{BChPTvalues}
\hskip-1cm\mbox{B$\chi$PT}:  && \al = (10.8\pm 0.7) \times 10^{-4} \mbox{ fm$^3$} , \nn\\
&&
\be = (4.0\pm 0.7) \times 10^{-4} \mbox{ fm$^3$},
\eea
and at the same time gives the red solid curves for
the cross section in \Figref{DRvsEFT}.
 The cyan-colored band around these curves corresponds to uncertainty of $\pm 0.7 \times 10^{-4}$ fm$^3$ in both $\al$ and $\be$, which is exactly the radius of
the red blob in  \Figref{potato}. The DR code of L'vov produces the dotted black curves for the same
values of $\al$ and $\be$, and the dashed blue curves for the PDG values given in \Eqref{PDGvalues}. The TAPS data are shown by black dots. The other data are properly referenced
in  \cite{Lensky:2009uv}.

The figure shows that for the same values of $\al$ and $\be$ the DR and $\chi$EFT yield
nearly the same results. Thus, the observed discrepancy between the $\chi$EFT
and PDG results 
is not because of the difference in theoretical approaches---they agree where they should, but because
the extraction from the TAPS data went bad apparently, at least in the error estimate. 
One can see with a bare eye that
the cross-section data do not prefer the blue-dashed curves (PDG value) over the black-dotted curves (B$\chi$PT value) to any significance.   In fact, at backward angles the B$\chi$PT
curves are even in a better agreement with the data. 

It is fair to conclude that the existing cross-section data do not corroborate the claimed
accuracy of proton polarizabilities. This is quite unfortunate because we do need
a precise knowledge of these quantities, whether it is to understand the atomic
measurements or to provide a basis for determinations of proton's spin polarizabilities.
A new accurate
 measurement of Compton scattering below pion-production threshold can of course
remedy this situation. It would be particularly interesting to use polarized beam, as  proposed at the HI$\ga$S facility \cite{Ahmed08}.
The polarized beam allows one to separate $\al$ and $\be$, which is very important since
the smallness of $\be$ makes it difficult for observation in the $\al$-dominated unpolarized
cross section. We have checked that the cross sections for 
the photon beam linearly polarized in the scattering plane ($||$) or orthogonal to it ($\perp$)
have the following expression at leading order in the low-energy expansion:
\beq
\begin{array}{l}
 \frac{d\si_{||}}{d\Omega}  -  \frac{d\si_{||}}{d\Omega}^{(\mathrm{Born})}  \hspace{-.1cm}  = 
- \frac{e^2}{2\pi M_p}\left( \frac{\nu'}{\nu} \right)^2  \nu\nu'     \big( \alpha \, z^2   +  \beta \, z \big), \nn\\
\frac{d\si_{\perp}}{d\Omega}  - \frac{d\si_{\perp}}{d\Omega}^{\!\!(\mathrm{Born})}   \hspace{-.1cm}
 =
-\frac{e^2}{2\pi M_p}\left( \frac{\nu'}{\nu} \right)^2  \nu\nu'     \big( \alpha    +  \beta \, z \big), \nn
\end{array}
\eqlab{polcs}
\eeq
and hence by measuring these cross sections one would be able to determine $\al$ and $\be$
independently.

\section{Radiative pion photoproduction: $\De$'s MDM}

Recent measurement of the radiative pion photoproduction ($\gamma p \to p  \pi^0 \gamma'$ )
\cite{Schumann:2010js} aimed at a determination of the $\De(1232)$-resonance
magnetic dipole moment (MDM).  The obtained data, however, could not be well-described using the existing theoretical frameworks, including a $\chi$EFT calculation 
\cite{Pascalutsa:2004je,Pascalutsa:2007wb}. Any extraction then becomes a moot point.

It is now better understood what the problem is. 
A large portion of the data lies outside 
the $\De$-resonance region. This is best seen in  by plotting the data as function
of the incoming photon energy,  see \Figref{Endep}, 
rather than as function of outgoing photon energy as
it's usually done. The red/black data points in this figure are from the dedicated experiment \cite{Schumann:2010js},
while the green/brown  points are from a preliminary analysis of newer data \cite{Prakhov10}.

The $\chi$EFT calculation clearly shows the resonant structure but the data-binning is
so wide that one cannot distinguish it. Of course, a comparison of theory with 
the data below or above the resonance should be applied with caution
since at least the $\chi$EFT calculation cannot be applied at arbitrary energies.
The present calculation \cite{Pascalutsa:2007wb}
is done at a next-to-leading order in the resonance regime
of the $\de$-counting scheme \cite{Pascalutsa:2003zk,Pascalutsa:2006up}, 
and as such works only in a 
narrow window around the resonance. This can be seen, e.g., from \Figref{piphoto} where the corresponding
description of pion-photoproduction observables is shown in the region of 50 MeV around the
resonance.

\begin{figure}
 \centerline{\epsfclipon  \epsfxsize=8cm %
  \epsffile{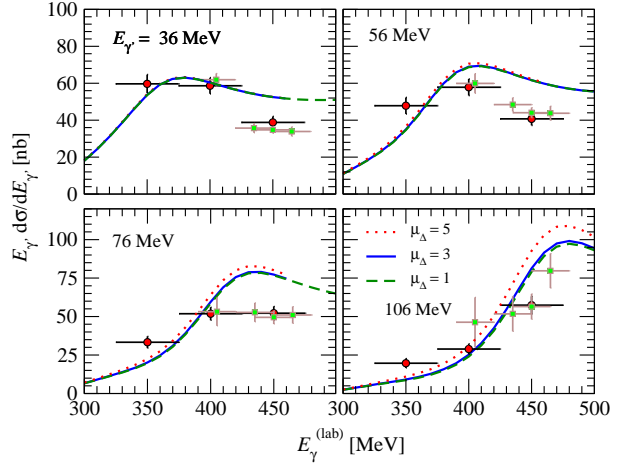} 
}  \caption{\small Differential cross section of  $\gamma p \to p  \pi^0 \gamma'$ as function
of photon beam energy. The curves are results   of $\chi$EFT calculation 
\cite{Pascalutsa:2007wb} for different values of the $\De$ MDM (in units of nuclear magneton). 
The data points are from Crystal Ball @MAMI Collaboration: red---\cite{Schumann:2010js},
green---\cite{Prakhov10}.}
\figlab{Endep}
\end{figure}

\begin{figure}
 \centerline{\epsfclipon  \epsfxsize=8cm%
  \epsffile{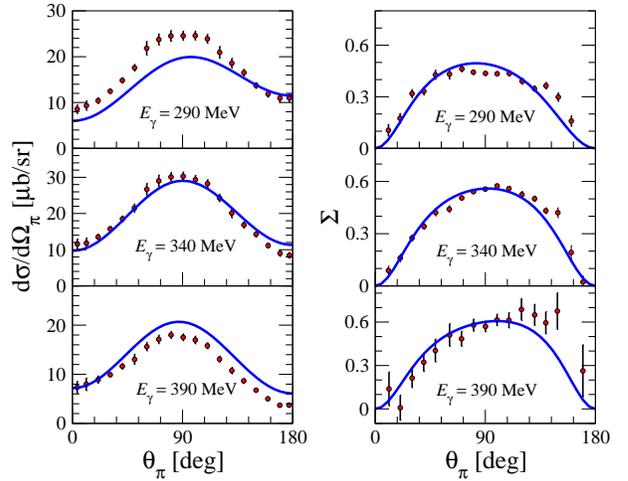} 
}
  \caption{\small Differential cross-section and beam asymmetry of $\ga p \to p \pi^0$
  in NLO $\chi$EFT \cite{Pascalutsa:2007wb,Pascalutsa:2006up}.}
  \figlab{piphoto}
\end{figure}

\begin{figure}
 \centerline{\epsfclipon  \epsfxsize=7cm%
  \epsffile{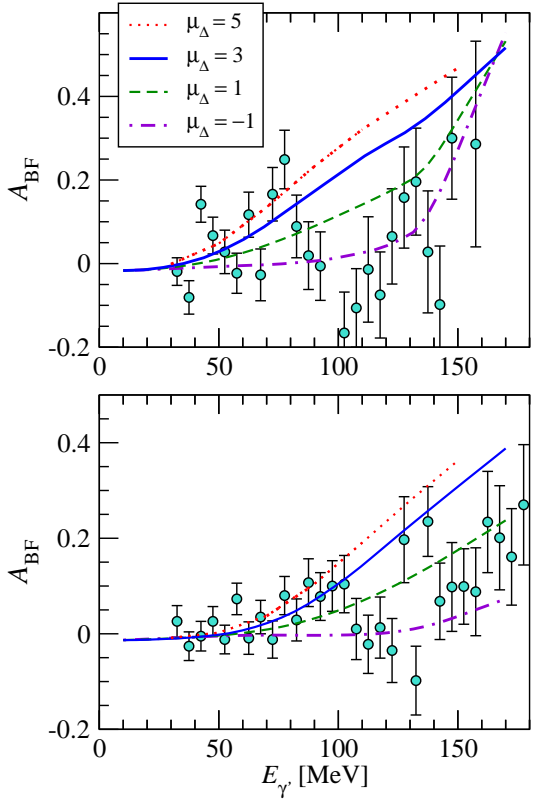} 
}
  \caption{\small Backward-forward asymmetry for different values of the $\De$ MDM, at fixed
  beam energy: 350 MeV -- upper panel, 400 MeV -- lower panel, as a function of outgoing photon energy. Data points are from a preliminary analysis of CB@MAMI Collaboration \cite{Schu10}.}
  \figlab{asym}
\end{figure}

The  \Figref{Endep} also shows that the differential cross section is largely insensitive
to the value of $\De^+$ MDM. In fact, the dependence of this observable on the MDM is
quadratic, i.e., an expansion of in powers of the outgoing photon energy ($E_\ga'$) 
can be written as:
\beq
E_\ga'\,\, d\si/ d E_{\ga}' =  s_0 + s_1\, E_{\ga}'+ s_2\,\mu_\De^2 \, E_{\ga}^{\prime 2} 
+ O(E_{\ga}^{\prime 3}),
\eeq
where real-valued $s_i$ depend on the mass, width, charge and spin of the resonance, but are independent of its magnetic moment. 

It is much easier and more feasible  to extract the MDM from a linearly-dependent quantity. One example of such an observable is a circularly-polarized photon beam asymmetry defined in \cite{Pascalutsa:2007wb}. Unfortunately this asymmetry is very small and tough to measure. 
Unpolarized observables of this kind are therefore welcome. 

We propose
an asymmetry  of the outgoing pion and photon landing 
 into different forward/backward hemispheres
versus landing the same forward/backward hemisphere. 
In terms of the 5-fold differential cross section
in the center-of-mass system this asymmetry is defined as:
 \bea
  A_{BF} &=& \frac{1}{d\si/dE_\ga'} 
 \Big[  \int_0^{\frac{\pi}{2}}\!\! d\Omega_{\ga\,'} \int_{\frac{\pi}{2}}^\pi\!\!  d\Omega_\pi 
 + \int_{\frac{\pi}{2}}^\pi \!\!  d\Omega_{\ga\,'}  \int_0^{\frac{\pi}{2}}\!\!  d\Omega_\pi \nn\\
 -&&  \hspace{-0.9cm}\int_0^{\frac{\pi}{2}}\!\! d\Omega_{\ga\,'}  \int_0^{\frac{\pi}{2}}\!\!  d\Omega_\pi 
 - \int_{\frac{\pi}{2}}^\pi \!\! d\Omega_{\ga\,'} \int_{\frac{\pi}{2}}^\pi\!\! d\Omega_\pi \Big] \frac{d^5 \si}{dE_\ga' d\Omega_{\ga\,'}  d\Omega_\pi}\, .\nn\\
\eea
Its low-energy
expansion goes as
\beq
A_{BF} =  a_0 + a_1\,\mu_\De \, E_{\ga}' 
+ O(E_{\ga}^{\prime 2}),
\eeq
and thus the MDM becomes seen at lower energy.
Some $\chi$EFT results for this asymmetry are shown in \Figref{asym}, and can be compared
to a preliminary analysis using the database of the dedicated CB@MAMI experiment \cite{Schumann:2010js}. One can see a significant sensitivity of this asymmetry on 
the MDM value. However, the present data are neither accurate nor consistent
enough to determine the
$\De^+$ MDM with any good precision.

\section{Conclusion}  
  The existing data on proton Compton scattering cross-sections do not corroborate
  the claimed (e.g., in PDG tables) precision of our knowledge of proton's electric and magnetic polarizabilities. New experiments, using polarized photon beam, are needed to correct this situation.
  
  A reliable extraction of the $\De$-resonance magnetic dipole moment from the process
  of radiative pion production will require a substantial improvement in the
  experimental database, as well as further systematic  refinements
   of the theoretical modeling.
\begin{theacknowledgments}
I am indebted to J\"urgen Ahrens for providing me with the results of L'vov's dispersion relations
shown in  \Figref{DRvsEFT}.
\end{theacknowledgments}





\end{document}